\documentclass[twocolumn]{aastex631}
\shorttitle{Atmosphere of Uranus}
\shortauthors{Fletcher}
\graphicspath{{./}{figures/}}
\begin{document}

\title{The Atmosphere of Uranus \\
(Oxford Research Encyclopedia of Planetary Science)  }

\correspondingauthor{Leigh N. Fletcher}
\email{leigh.fletcher@le.ac.uk}

\author[0000-0001-5834-9588]{Leigh N. Fletcher}
\affiliation{School of Physics and Astronomy, University of Leicester, University Road, Leicester, LE1 7RH, United Kingdom.}

%% Mark off the abstract in the ``abstract'' environment. 
\begin{abstract}

%The first lesson in the tutorial is to remind authors that the AAS Journals, the Astrophysical Journal (ApJ), the Astrophysical Journal Letters (ApJL), the Astronomical Journal (AJ), and the Planetary Science Journal (PSJ) all have a 250 word limit for the  abstract

% Article Summary
%The article summary should be a brief synopsis of the topic, not of the article itself. The summary should be roughly equivalent to a definition, one or two paragraphs in length. Unlike a traditional “abstract,” the summary should be able to stand on its own as a useful piece of content without reference to a larger article. Please also note that our website platform does not support citations in the summary. The summary will publish right away and serve as a preview for the full article. When the full article is published, the summary will appear at the beginning. If you would like to make changes to your summary when you submit the final article, please include a revised copy.

Uranus provides a unique laboratory to test our understanding of planetary atmospheres under extreme conditions.  Multi-spectral observations from Voyager, ground-based observatories, and space telescopes have revealed a delicately banded atmosphere punctuated by storms, waves, and dark vortices, evolving slowly under the seasonal influence of Uranus' extreme axial tilt.  Condensables like methane and hydrogen sulphide play a crucial role in shaping circulation, clouds, and storm phenomena via latent heat release through condensation, strong equator-to-pole gradients suggestive of equatorial upwelling and polar subsidence, and through forming stabilising layers that may decouple different circulation and convective regimes as a function of depth.  Phase transitions in the watery depths may also decouple Uranus' atmosphere from motions within the interior.  Weak vertical mixing and low atmospheric temperatures associated with Uranus' negligible internal heat means that stratospheric methane photochemistry occurs in a unique high-pressure regime, decoupled from the influx of external oxygen.  The low homopause also allows for the formation of an extensive ionosphere.  Finally, the atmosphere provides a window on the bulk composition of Uranus - the ice-to-rock ratio, supersolar elemental and isotopic enrichments inferred from remote sensing and future \textit{in situ} measurements - providing key insights into its formation and subsequent migration. 

This review reveals the state of our knowledge of the time-variable circulation, composition, meteorology, chemistry, and clouds on this enigmatic `Ice Giant,' summarising insights from more than three decades of observations, and highlighting key questions for the next generation of planetary missions.  As a cold, hydrogen-dominated, intermediate-sized, slowly-rotating, and chemically-enriched world, Uranus could be our closest and best example of atmospheric processes on a class of worlds that may dominate the census of planets beyond our own Solar System. 

\end{abstract}

%% Keywords should appear after the \end{abstract} command. 
%% The AAS Journals now uses Unified Astronomy Thesaurus concepts:
%% https://astrothesaurus.org
%% You will be asked to selected these concepts during the submission process
%% but this old "keyword" functionality is maintained in case authors want
%% to include these concepts in their preprints.

%% From the front matter, we move on to the body of the paper.
%% Sections are demarcated by \section and \subsection, respectively.
%% Observe the use of the LaTeX \label
%% command after the \subsection to give a symbolic KEY to the
%% subsection for cross-referencing in a \ref command.
%% You can use LaTeX's \ref and \label commands to keep track of
%% cross-references to sections, equations, tables, and figures.
%% That way, if you change the order of any elements, LaTeX will
%% automatically renumber them.
%%
%% We recommend that authors also use the natbib \citep
%% and \citet commands to identify citations.  The citations are
%% tied to the reference list via symbolic KEYs. The KEY corresponds
%% to the KEY in the \bibitem in the reference list below. 

\tableofcontents

%Your article should be 6,000-10,000 words in length, including Summary, Keywords, Main Essay, and References. 

\section{Introduction} 
\label{sec:intro}

% Introductory Paragraphs (400 – 500 words)
%  Define the topic you will cover and why.
%  Outline the areas of science that inform your work. Note how this work fits in the larger context of planetary science.

Of all the primary planets in our Solar System, Uranus could be considered as the most unusual.  Knocked onto its side, potentially by a cataclysmic impact during the epoch of planetary formation and migration \citep{66safronov, 86stevenson, 19kegerreis}, the $98^\circ$ axial tilt subjects the atmosphere to extreme seasons, with each pole spending four decades in the darkness of polar winter, and four decades in perpetual sunlight.  Unlike the other giants, Uranus lacks any detectable internal heat source emanating from its deep interior \citep[i.e., no appreciable `self luminosity,'][]{90pearl}, which may be related to the apparent dearth of large-scale meteorological activity and atmospheric mixing.  The sluggish vertical mixing results in stratospheric chemistry operating in a higher-pressure regime quite unlike that found on other worlds.  Compared to the Gas Giants Jupiter and Saturn, Uranus (and Neptune) possess substantial super-solar enrichments in heavy elements, such that the resulting volatiles (methane, hydrogen sulphide, ammonia, and water) play a crucial role in the stability and energetics of the weather layer.  These gaseous species exhibit strong equator-to-pole gradients, hinting at large-scale tropospheric circulation.  Combined with the smaller size (4.2 Earth radii), slower rotation (17 hours), and colder tropospheric temperatures ($\sim50$ K, sufficient for methane to condense) compared to the Gas Giants, these factors all have stark implications for Uranus' banded structure, circulation, meteorology, and photochemistry.  Indeed, Uranus' atmosphere occupies a unique region of dynamical and chemical parameter space not found elsewhere in our Solar System.

Uranus therefore provides a planetary-scale laboratory to test our understanding of atmospheric physics and chemistry under extreme environmental conditions.  Comparison to Neptune, the archetype for a seasonal Ice Giant, reveals how planets can evolve down different evolutionary paths despite their common origins, allowing us to contrast the extreme seasons and negligible internal heat flux of tilted Uranus with the ever-changing storms of Neptune.  Taken together, the Ice Giants are our closest and best examples of a class of `intermediate-sized' astrophysical object that may be common in our universe.  Indeed, they are only slightly larger than the sub-Neptunes that dominate the current exoplanetary census \citep{18fulton}.  Exploration of Uranus and Neptune therefore reveals how atmospheric circulation, planetary banding, and seasonal photochemistry operate on cold, intermediate-sized, chemically-enriched, and slow-rotating hydrogen-rich worlds \citep{19fletcher_V2050, 20wakeford}.  Finally, the basic composition of Uranus - its ice-to-rock ratio, supersolar elemental enrichment, and isotopic composition - provides an essential missing piece in the puzzle of the architecture of our Solar System, revealing the location and timescale for Uranus' accretion \citep{18mousis}, and providing insights into its subsequent migration and thermal evolution.

More than 35 years have now passed since the only spacecraft to visit Uranus, Voyager 2, flew past in 1986 on its journey out of our Solar System.  Our knowledge of Ice Giant atmospheres was reviewed shortly after the Voyager encounters by \citet{93lunine}.  Since then, observations from ground- and space-based observatories have revealed a picture of Uranus' atmosphere that is complex, perplexing, and altogether unlike that seen on the Gas Giants.   The reader is directed to several recent reviews that investigate aspects of Uranus' atmosphere in great depth:  the bulk composition and implications for planetary origins \citep{18mousis, 20atreya_icegiant}; the meteorology \citep{19hueso, 20hueso}, global circulation \citep{20fletcher_icegiant}, and electrical influences \citep{20aplin} on Uranus' troposphere; stratospheric chemistry \citep{20moses}; and upper atmospheric structure \citep{20melin}. Looking to the future, Uranus' atmosphere will be a key target for future orbital and \textit{in situ} exploration, alongside in depth studies of the interior, magnetosphere, rings and icy satellites \citep{12arridge, 18mousis, 19hofstadter, 19fletcher_V2050}. 

This article attempts to create a synthesis of these works to review the current state of our understanding of this enigmatic world.  Section \ref{partone} provides a brief observational history of Uranus, and describes the basic properties of its atmosphere.  Section \ref{parttwo} then explores our knowledge of Uranus' circulation, meteorology, and chemistry, revealing how the atmosphere may be connected to both the deep, hidden, water-rich interior, and the charged environment of the magnetosphere.  We will conclude in Section \ref{conclude} with the significant outstanding questions that must be addressed by future missions to this extreme Ice Giant.

\section{Observations of Uranus}
\label{partone}

% Part One (2000 – 3000 words)
%  Chart our understanding of the topic as it has developed over time: consider when and how the topic appeared and then took on its current form.
%  Provide balanced coverage of the context, the controversies, and the debates that have informed and helped to form the topic, and that animate it now.
%  Discuss foundational and notable discoveries or advances and those who made them within their context and current perspectives; include biographical details as needed.

\subsection{A Brief Observational History}

\begin{figure*}
\begin{center}
\includegraphics[angle=0,width=1.0\textwidth]{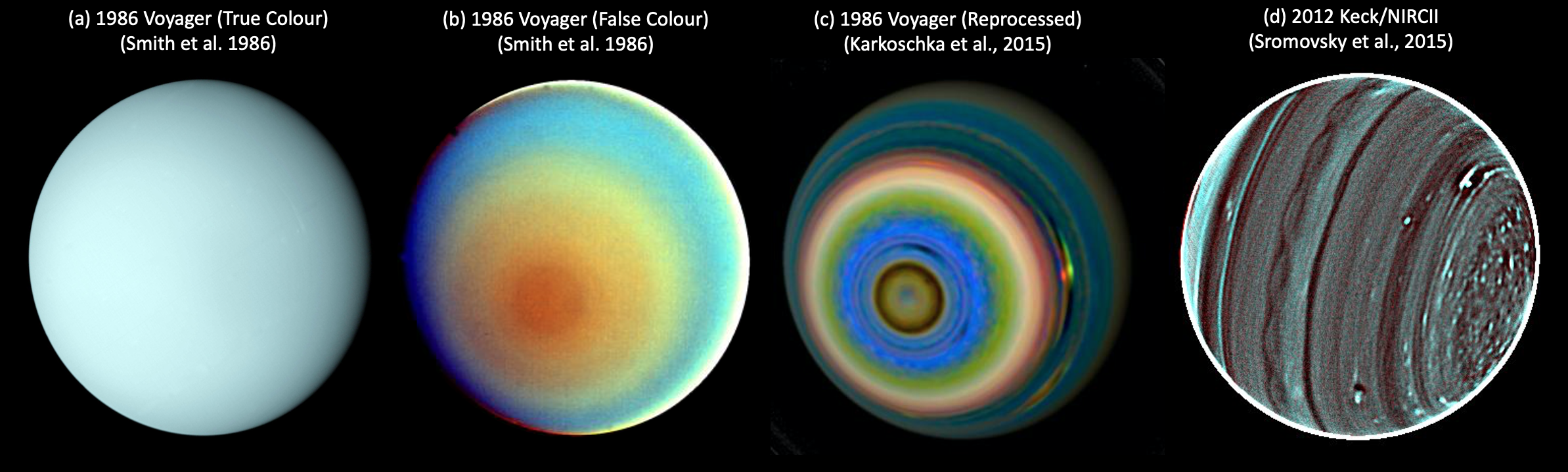}
\caption{Evolution of Uranus observations from Voyager (a-c) to ground-based observatories (d).  Uranus originally appeared rather featureless in Voyager-2 observations \citep{86smith} despite attempts to stretch the contrast (b). Modern image processing techniques (c) reveal more structure \citep{15karkoschka}, but longer-wavelength observations and derotation techniques \citep[e.g., Keck H-band images in 2012,][]{15sromovsky} are one of the best ways to investigate the details of the atmosphere.}
\label{ura_voyager}
\end{center}
\end{figure*}

Although the colourful cloud patterns of Jupiter and Saturn had been intensely studied for centuries, Uranus' atmosphere remained stubbornly out of reach.  Faint dusky belts were suggested as early as 1883 by Young, and 1884 by the Henrys \citep{1883young, 1884henry}, and can be seen in drawings \citep{65alexander} by G. Fournier (1913) and R.L. Waterfield (1915-16).  As Uranus was at autumn equinox (planetocentric solar longitude, $L_s=180^\circ$) in 1882, and spring equinox ($L_s=0^\circ$) in 1923, both hemispheres would have been visible at these times, but the curious orientations of the sketched bands left their existence in doubt.  Using spectroscopy, \citet{1904slipher} discovered dark bands in the visible spectrum which were later demonstrated to be signatures of methane \citep{1932wildt}.  Furthermore, spectral features identified by \citet{49kuiper} were found to be due to molecular hydrogen \citep{52herzberg}, such that the basic atmospheric composition (a hydrogen atmosphere enriched in methane) was known long before the space age.

\begin{figure*}
\begin{center}
\includegraphics[angle=0,width=1.0\textwidth]{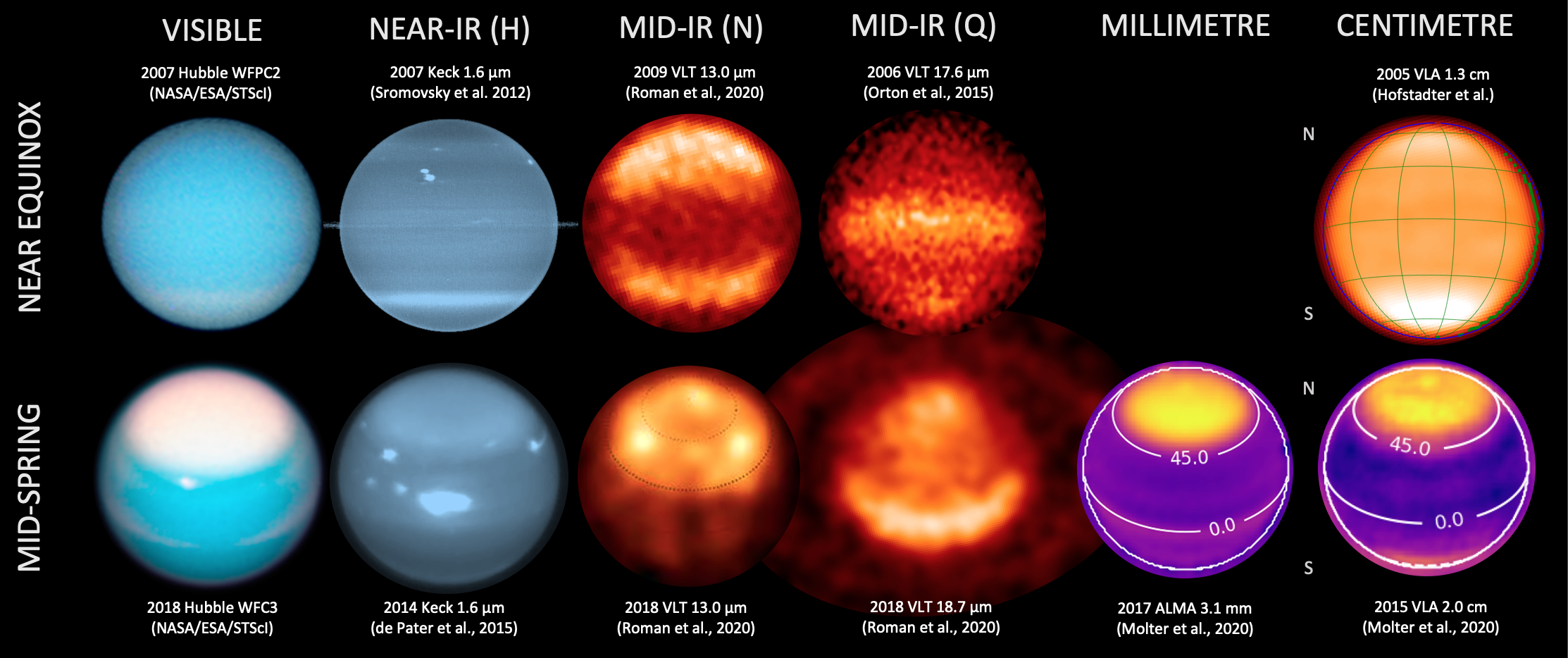}
\caption{Imaging of Uranus from the visible to the radio, in two groups:  near the 2007 equinox (top row), and during mid-northern spring (bottom row).  From left to right, we show visible-light images from Hubble WFPC2 and WFC3 (Credit: NASA/ESA/STScI); near-infrared images from Keck \citep{15sromovsky, 15depater}; mid-IR images of the stratosphere (13 $\mu$m) and troposphere (18.7 $\mu$m) from VLT \citep{20roman}; millimetre observations from ALMA \citep{20molter}; and centimetre-wave observations from VLA \citep{20molter}.  Uranus has been oriented so the the north pole is at the top.}
\label{ura_montage}
\end{center}
\end{figure*}

The secrets of Uranus' atmosphere only really started to be revealed by Voyager 2 encounter on 1986-01-24 as part of its `Grand Tour,' and this 1970s spacecraft remains the only mission to ever provide close up views \citep{86smith, 87lindal, 86tyler}.  At the time of the flyby, Uranus was just past northern winter solstice ($L_s=271.4^\circ$), meaning that the southern polar region was pointing almost directly at the Sun.  Visible light images from the Voyager cameras \citep{86smith} in Fig. \ref{ura_voyager} revealed a greeny-blue world due to the high abundance of red-absorbing methane, with faint banding associated with methane condensation clouds, but very few discrete features - only eight features were available to understand the pattern of zonal winds.  Early Hubble observations of Uranus in 1994-1995, during northern winter ($L_s=310^\circ$) \citep{97karkoschka} were similarly bland, although near-IR images began to provide more contrast in 1997 ($L_s=319^\circ$) \citep{98karkoschka, 01karkoschka_rings, 01karkoschka}.  Modern image processing of the Voyager-2 data by \citet{15karkoschka} revealed many more southern hemisphere features (Fig. \ref{ura_voyager}c), but the perception of Uranus as a featureless planet had already started to take root, albeit unfairly.  Subsequent Earth-based near-IR imaging (Fig. \ref{ura_voyager}d) exploited strong methane and hydrogen absorption bands in the 1-2.2 $\mu$m range (`red wavelengths') to improve contrast, sensing the aerosol distribution (condensate clouds and photochemical hazes) as a function of altitude \citep[e.g.,][and others]{12fry, 15sromovsky}.  Uranus passed northern spring equinox ($L_s=0^\circ$) in December 2007 (Fig. \ref{ura_montage}), permitting the first views of the north pole as it emerged into sunlight \citep{09sromovsky, 11depater, 12sromovsky}.  Today, modern equipment and image processing techniques now allow amateur observers to track prominent storms and bands on Uranus \citep{15depater}.  These visible and near-IR observations are summarised by \citet{19hueso} and described in Section \ref{parttwo}, and started to reveal Uranus' finely banded structure, bright clouds, vortices, and ephemeral storms in exquisite detail (Fig. \ref{ura_montage}).  

\begin{figure}
\begin{center}
\includegraphics[angle=0,height=0.8\textheight]{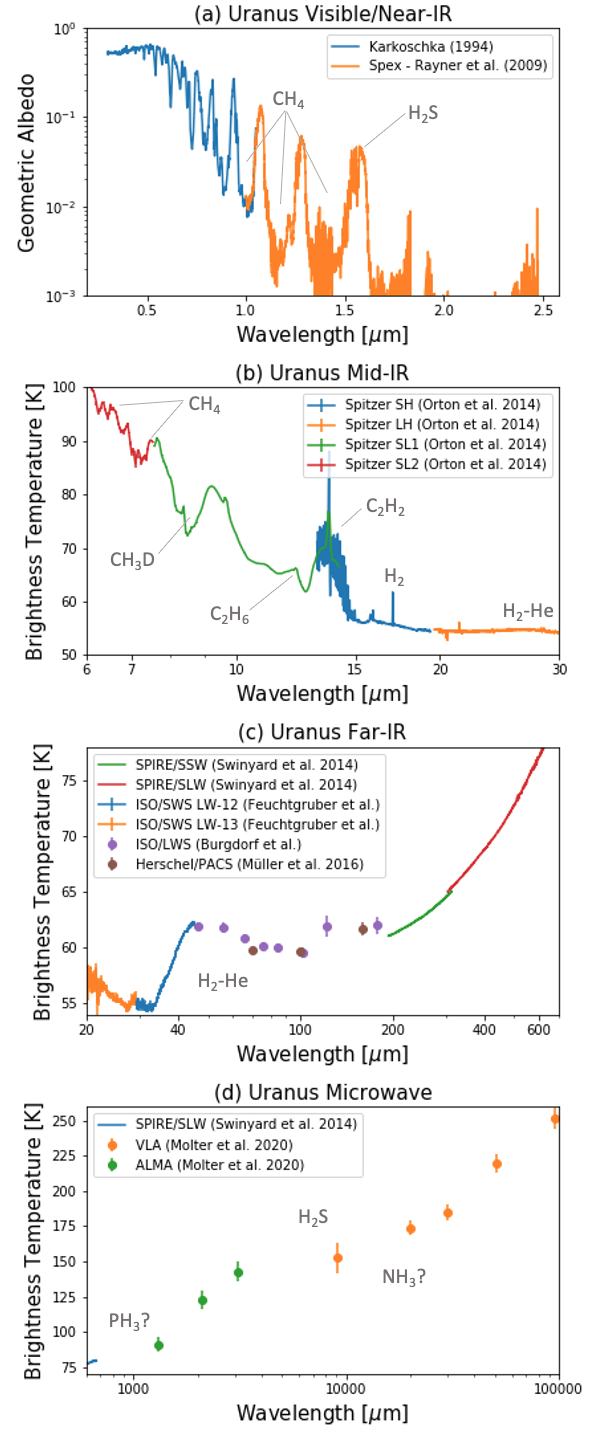}
\caption{Disc-integrated spectroscopy of Uranus from the visible to the radio, with key spectroscopic features labelled.  Panel (a) shows reflected sunlight measured by \citet{94karkoschka} and \citet{09rayner}.  Panel (b) presents Spizter/IRS observations of Uranus in low and high-resolution modes \citep{14orton}.  Panel (c) reveals the far-IR continuum dominated by collision-induced H$_2$-He continuum, constructed from Herschel SPIRE shortwave and longwave spectra \citep{14swinyard} and PACS photometry \citep{16muller}, along with ISO Shortwave (courtesy of H. Feuchtgruber) and Longwave (courtesy of M. Burgdorf) spectra.  Panel (d) shows VLA and ALMA observations from \citet{20molter}.  Panels (b-d) are presented in brightness temperatures as a proxy for the atmospheric temperatures in the line-formation regions.} 
\label{ura_spectra}
\end{center}
\end{figure}

Interpreting aerosol distributions requires an understanding of atmospheric temperature and composition, which requires observations in the mid-IR, far-IR, sub-millimetre and radio (Fig. \ref{ura_montage}).  The extreme cold and distance has severely limited our ability to explore this wavelength domain.  Although Voyager 2 carried a capable thermal-IR spectrometer, the radiance was so low that only the hydrogen-helium continuum in the 25-50 $\mu$m range could be used, revealing tropospheric temperature contrasts between cold mid-latitudes, associated with upwelling, and the warm equator and pole, potentially due to subsidence \citep{98conrath}. Space telescopes like ISO \citep{00encrenaz}, Herschel \citep{13feuchtgruber} and Spitzer \citep{14orton} provided high-quality spectra from the infrared to the sub-mm (Fig. \ref{ura_spectra}), but these were disc-integrated, unable to resolve contrasts across Uranus' atmosphere.  The development of ground-based observatories with $>8$-m primary mirrors (Keck, VLT, Gemini, and Subaru) now permits spatially-resolved imaging of Uranus' thermal emission in the 7-25 $\mu$m range, sensing both tropospheric (0.1-1.0 bar) and stratospheric (0.1-10 mbar) temperatures \citep{15orton, 20roman}, albeit with a low signal-to-noise as shown in Fig. \ref{ura_montage}.  Such temperature measurements are supported by occultations of stars in the UV \citep{90bishop}.  At longer wavelengths, centimetre and millimetre-wavelength arrays like the Karl G. Jansky Very Large Array (VLA) and Atacama Large Millimeter Array (ALMA) \citep{88depater, 03hofstadter, 14depater} provide maps of gaseous contrasts at greater depths (Fig. \ref{ura_montage}), with spatial resolutions now approaching those of the reflected-sunlight images \citep{18depater, 20molter}.  These microwave observations sense thermal emission down to tens of bars, modulated by the pressure-broadened wings of NH$_3$ and H$_2$S, along with other potential contributions from CH$_4$, CO, H$_2$O, and PH$_3$, although uniquely distinguishing between these contributions remains a challenge \citep[see (Fig. \ref{ura_spectra}d,)][]{91depater_radio}.  Finally, observations of thermospheric emission from H$_2$ and H$_3^+$ also enable studies of the temperatures and circulation of the upper atmosphere (ionosphere and thermosphere).  

Taken together, the observations of reflected sunlight and thermal emission in Fig. \ref{ura_spectra} are revealing Uranus' atmosphere in three dimensions, and studying how it evolves over time, as discussed in the following sections.

\subsection{Vertical Structure and Composition}

\begin{figure}
\begin{center}
\includegraphics[angle=0,width=0.45\textwidth]{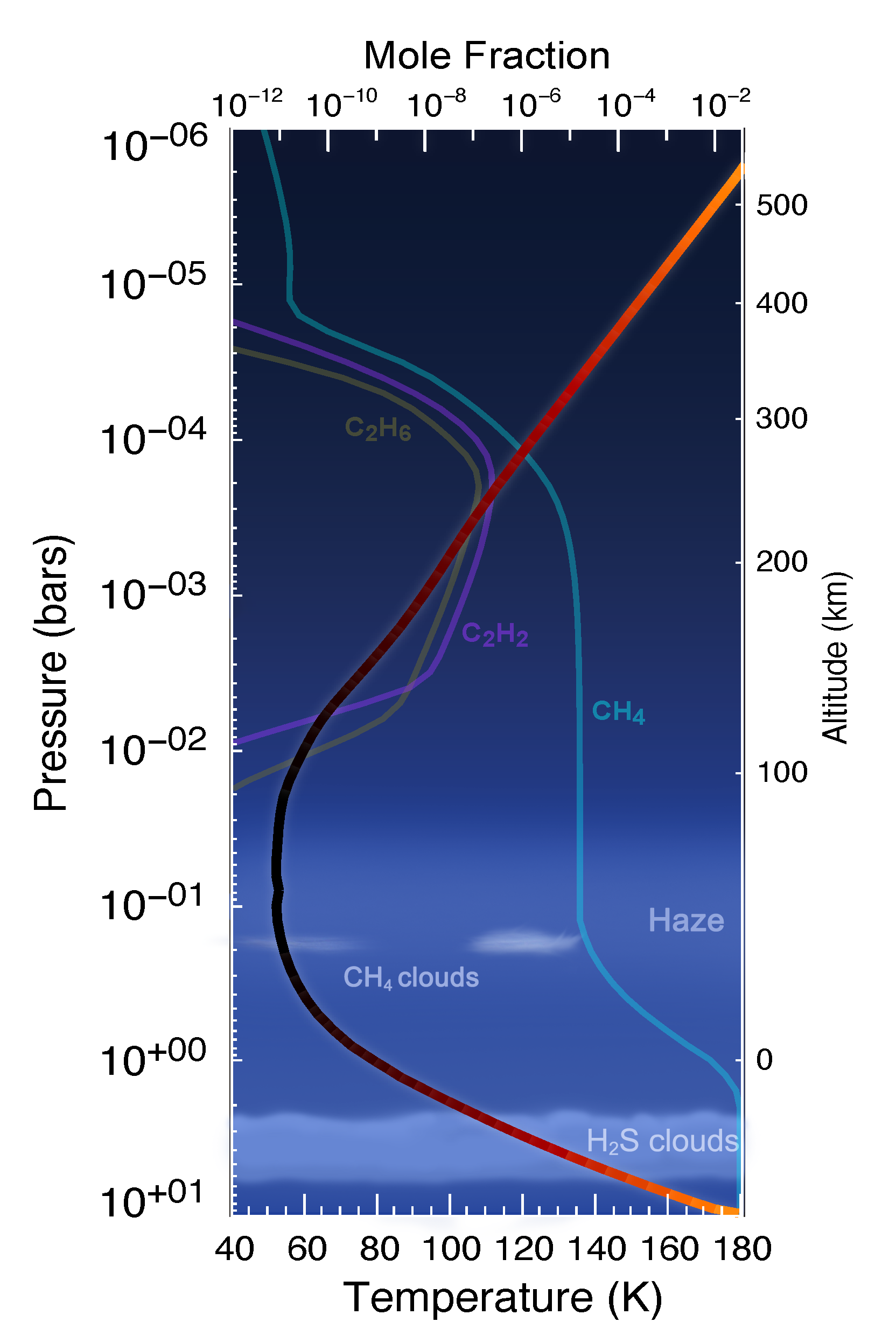}
\caption{Uranus' temperature structure derived from Spitzer observations \citep{14orton}, compared to the vertical profiles of methane, ethane, and acetylene \citep{18moses} and the approximate locations of Uranus' CH$_4$ and H$_2$S clouds.  Credit: M.T. Roman.} 
\label{temp-clouds}
\end{center}
\end{figure}

\begin{figure}
\begin{center}
\includegraphics[angle=0,width=0.4\textwidth]{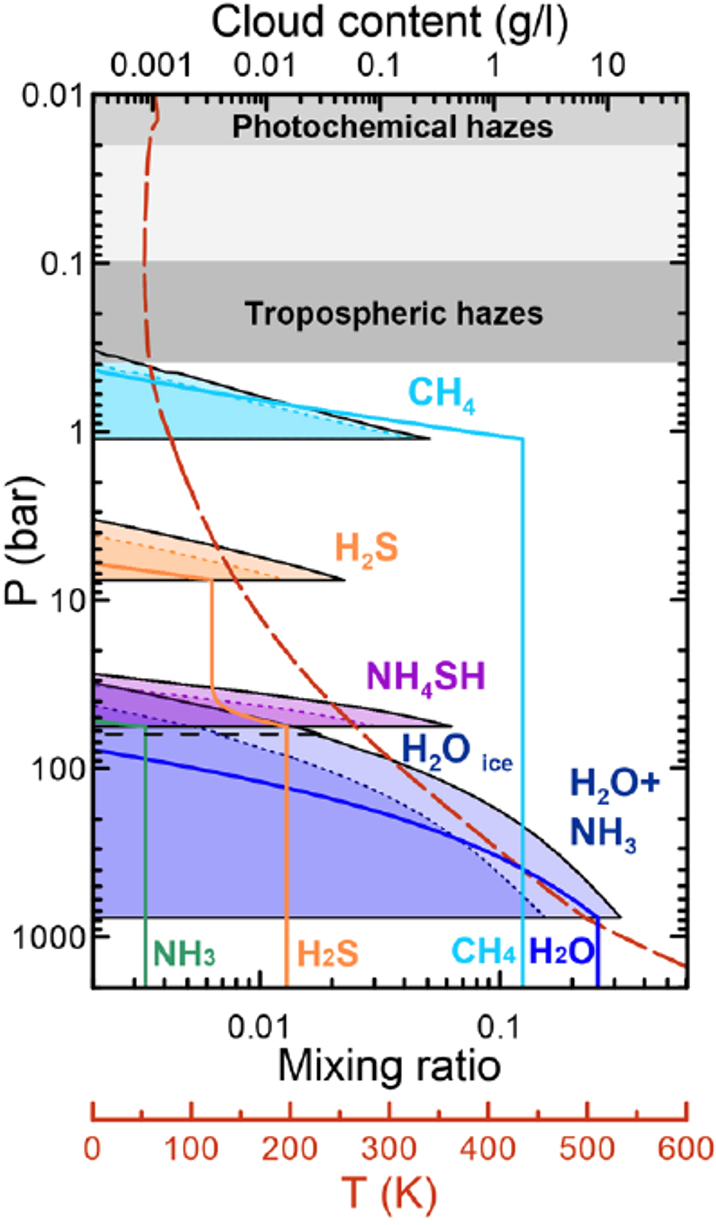}
\caption{Approximate locations of Uranus' clouds from thermochemical equilibrium condensation \citep{19hueso}.  Mixing ratios of volatiles (lower axis) are for a $30\times$ solar enrichment \citep[approximately consistent with the $37^{+13}_{-6}$ enrichment reported by][]{20molter}, except for NH$_3$ which is assumed to be depleted ($3\times$ solar).  A moist adiabatic temperature profile, extrapolated from Voyager radio occultation profiles, is shown as the dashed red-brown line.  Cloud densities (black continuous lines) are referenced to the top axis - see Fig. 9 of \citet{19hueso} for full details.}
\label{clouds}
\end{center}
\end{figure}

% Methane
Before delving into the details of the processes shaping Uranus' atmosphere, we first review the basic properties of the troposphere and stratosphere, shown in Fig. \ref{temp-clouds}.  Based on Uranus' expected formation region beyond the snowline, the bulk of the planet is thought to be made up of 10-20\% hydrogen and helium, plus 80-90\% heavier elements \citep{95hubbard, 19podolak}. The proportion of rocky refractory materials versus ice-rich materials remains unclear, but the most abundant condensable species are expected to be CH$_4$, H$_2$O, NH$_3$, and H$_2$S.  Of these only CH$_4$ and H$_2$S \citep{18irwin_h2s} have been directly measured:  methane was first estimated to comprise around 1.6-2.3\% from Voyager radio occultations \citep{87lindal} and ground-based visible quadrupole measurements \citep{95baines}, but later estimates are in the 2.7-3.5\% range \citep{09karkoschka,19sromovsky}.  Based on these chemical abundances, and the Clausius-Clapeyron equation \citep{73weidenschilling, 04sanchez_clouds}, Uranus' topmost clouds (Fig. \ref{clouds}) are expected to be comprised of a thin layer of CH$_4$ ice with a base near 1.3 bars and approximately 80 K, where latent heat released by condensation was observed to modify the temperature lapse rate observed in Voyager radio occultations \citep{87lindal}.  Cloud sounding via visible and near-IR spectroscopy (Fig. \ref{ura_spectra}a) confirms the presence of optically-thin hazes in the upper troposphere and stratosphere, above a thin methane ice cloud near 1.3 bars, and thicker H$_2$S cloud in the 2-4 bar range \citep{09irwin_ura, 13tice, 15dekleer}.

% \begin{figure*}
% \begin{center}
% \includegraphics[angle=0,width=1.0\textwidth]{}
% \caption{Vertical structure of Uranus' temperatures and clouds.}
% \label{temp-clouds}
% \end{center}
% \end{figure*}

% Helium
H$_2$ and He become more enriched with height as the volatile gases condense out to form clouds, such that the observable upper troposphere is more `dilute' and lacks signatures of H$_2$O and NH$_3$ and is unlikely to be representative of the bulk.  The helium mass fraction $Y=0.262\pm0.048$ was reported by \citet{87conrath} using Voyager data, which is just slightly less than the protosolar value (0.278) \citep{09lodders}.

% Nitrogen and Sulphur
The methane deep volume mixing ratios of \citep{19sromovsky} suggest a bulk carbon enrichment of $\sim50-85\times$ the protosolar abundance \citep{09asplund}, which is some ten times larger than that seen on the Gas Giants. However, there remains the possibility that the elements are not uniformly enriched.  Radio observations \citep{78gulkis, 91depater_radio} suggest a significant superabundance of H$_2$S ($37^{+13}_{-6}\times$ protosolar) compared to NH$_3$ ($1.4^{+0.5}_{-0.3}\times$ solar) \citep{20molter}, which is quite different to Jupiter's approximately solar N/S ratio, and the low abundance of sulphur in a solar-composition mixture.  We expect NH$_3$ and H$_2$S to combine to form a solid NH$_4$SH cloud layer in the 30-50 bar region \citep{73weidenschilling, 04sanchez_clouds, 05atreya}, leaving only the most abundant species - H$_2$S - at lower pressures \citep{94deboer}, as shown in Fig. \ref{clouds}.  The superabundance of H$_2$S has been supported by measurements of microwave opacity of H$_2$S in the lab \citep{96deboer}, and by the direct detection of H$_2$S vapour above the clouds in the H-band (Fig. \ref{ura_spectra}a) near 1.57-1.58 $\mu$m \citep{18irwin_h2s}.  Condensation of H$_2$S ice therefore forms an important cloud deck near 2-4 bars.  But does this sulphur excess imply a formation scenario where H$_2$S accretion was more efficient, potentially via trapping in clathrates \citep{04hersant}, or are we being misled because of chemistry in the deeper layers, where some NH$_3$ and H$_2$S will be dissolved in a deep water (solution) cloud, a liquid water ocean at tens of kilobars, and potentially an ionic/superionic ocean and hundreds of kilobars \citep{20atreya_icegiant}?  Even sampling the well-mixed abundances of NH$_3$ and H$_2$S below the NH$_4$SH cloud will be challenging, given that NH$_3$ is not well-mixed beneath the foreseen cloud base on Jupiter \citep{17bolton}. 

% Deep O/H
Observationally, we find ourselves using tropospheric measurements in the CH$_4$ ($\sim1.3$ bar) and H$_2$S (2-4 bar) cloud regions to infer the properties of Uranus' deeper interior.  The upper troposphere is effectively desiccated of other volatiles, so accessing Uranus' bulk water abundance, and thus the O/H ratio, remains a significant challenge.  Both planetary interior models \citep{95guillot} and measurements of CO \citep{17cavalie, 20venot} suggest that the H$_2$O abundance is potentially much larger, placing the water cloud base at hundreds of bars in Fig. \ref{clouds}.  Not only is this inaccessible to an entry probe, but it is also a challenge for microwave sounding due to our lack of knowledge of the deep temperature structure, as measured thermal profiles from Voyager end at 2.3 bar \citep{87lindal}.  Indeed, the potentially water-rich layers would cause lapse rates to depart substantially from adiabatic behaviour \citep{17leconte}, lending further complexity to Uranus' deep temperature profile assumed in Fig. \ref{clouds}.

% Disequilibrium species
Estimating Uranus' water content therefore relies on indirect approaches, such as using the chemistry of disequilibrium species.  Thermochemical equilibrium controls composition in the deeper atmosphere \citep{86fegley}, but when the rate of vertical transport is faster than the rate of chemical destruction, the composition can be frozen in at abundances representative of the `quench' point.  Thus disequilibrium species can be studied at higher, colder altitudes, but none of the potential species studied by \citet{86fegley} have been definitively detected.  CO, which is present in the stratosphere from external sources \citep{13cavalie_ura}, has only an upper limit in the troposphere \citep{13teanby}, so \citep{20teanby} point out that there is no need to bring CO upwards from an oxygen-rich interior to explain the measurements.  Nevertheless, if CO is present, then \citet{17cavalie} estimated that O/H is $<160\times$ the solar value in Uranus, but these have been revised downwards to $<45\times$ solar \citep{20cavalie} using a revised chemical and transport scheme \citep{20venot}, highlighting the potential model-dependency associated with this approach. Vertical transport, parameterised through an eddy diffusion coefficient in these chemical schemes, may also depend on latitude, further confusing the picture.  Ethane \citep{94fegley} and PH$_3$ could also be used to constrain O/H, but neither have been detected, and only upper limits are available for PH$_3$ from millimetre wavelengths \citep{09moreno_dps, 19teanby}.  And just as CO could aid in measuring the O/H abundance; molecular N$_2$ could aid in the deep N/H abundance:  N$_2$ contributes to the molecular weight and collision-induced continua in the infrared, and while this has been studied on Neptune \citep{93conrath_nep}, it has not been assessed on Uranus, in part because of the expected weak vertical mixing. Uranus' bulk O/H, N/H and C/O ratios therefore remain extremely uncertain.

% Nobel gases
Further insights into the bulk composition come from isotopic ratios.  The supersolar D/H ratio in H$_2$ derived from Herschel measurements \citep{13feuchtgruber} suggests accretion of a substantial water-ice-rich component, perhaps in addition to CO-rich ices \citep{14alidib} and rocky materials.  Furthermore, if Uranus accreted from amorphous ices then each element (C, N, S, and the noble gases Ar, Kr, Xe) should show similar enrichments with respect to protosolar, as the trapping efficiencies in amorphous ices are similar at low temperatures \citep{99owen}. Alternatively, if the volatiles were originally trapped in clathrates, then the trapping efficiency can vary substantially from molecule to molecule \citep{18mousis}.  Direct or indirect measurements of the elemental abundances in Uranus' atmosphere are therefore needed to understand which primordial icy and rocky reservoirs contributed to Uranus during its formation \citep{18mousis}.

Having discussed the basic composition and vertical structure of Uranus' atmosphere, Section \ref{parttwo} explores the various physical and chemical processes that reshape the temperatures, winds, clouds, and composition as a function of time and location.

\section{Atmospheric Processes}
\label{parttwo}

% Part Two (2500 – 3500 words)
%  Present the current state of the science, discipline or areas of study that your article focuses on, including strengths and weaknesses. Include observational, theoretical, and experimental techniques used.
%  Refer to work in as many other countries as is sensible.
%  You may add material from your own research in moderation.

\subsection{Atmospheric Circulation}
\label{circulation}

\subsubsection{Planetary Banding}
The zonal, meridional, and vertical motions of Uranus' troposphere and stratosphere may be studied via observations of zonal winds, temperature, composition, and aerosols, and how these parameters change over time.  Contrast-enhanced mages in reflected sunlight, such as those in Fig. \ref{ura_voyager}-\ref{ura_montage} \citep{12fry, 15karkoschka, 15sromovsky} reveal a banded atmosphere, with fine-scale $5-20^\circ$ latitudinal lanes of different reflectivity.  These contrasts are probably due to variations in aerosol optical depth with latitude, although \citet{15karkoschka} suggest that at least one of the bands could have been caused by variations in aerosol absorption, suggestive of different materials in some bands.  Unlike Jupiter and Saturn, this banding appears to have no strong correlation with the zonal winds and temperatures (see Section \ref{winds-temp}), which means that we cannot define canonical Ice Giant `belts' and `zones' in quite the same way \citep{20fletcher_icegiant}.  Furthermore, zonal medians of the reflectivity maps \citep{15sromovsky} indicate that these zonal contrasts are not static, but change from observation to observation, potentially due to obscuration of the banded structure by discrete features.  However, millimetre (1.3-3.1 mm) observations of Uranus from ALMA in 2017-2018, and centimetre (0.9-10 cm) observations from VLA in 2015 \citep[Fig. \ref{ura_montage},][]{20molter} suggest subtle thermal banding in the 1-50 bar range that is qualitatively similar to that observed in reflected sunlight in the 1-4 bar range.  This suggests that Uranus' albedo contrasts could be related to thermal and/or compositional contrasts in the troposphere, although the long-term stability and location of the zonal bands remains to be rigorously assessed.

\subsubsection{Winds and Temperatures}
\label{winds-temp}

\begin{figure}
\begin{center}
\includegraphics[angle=0,width=0.5\textwidth]{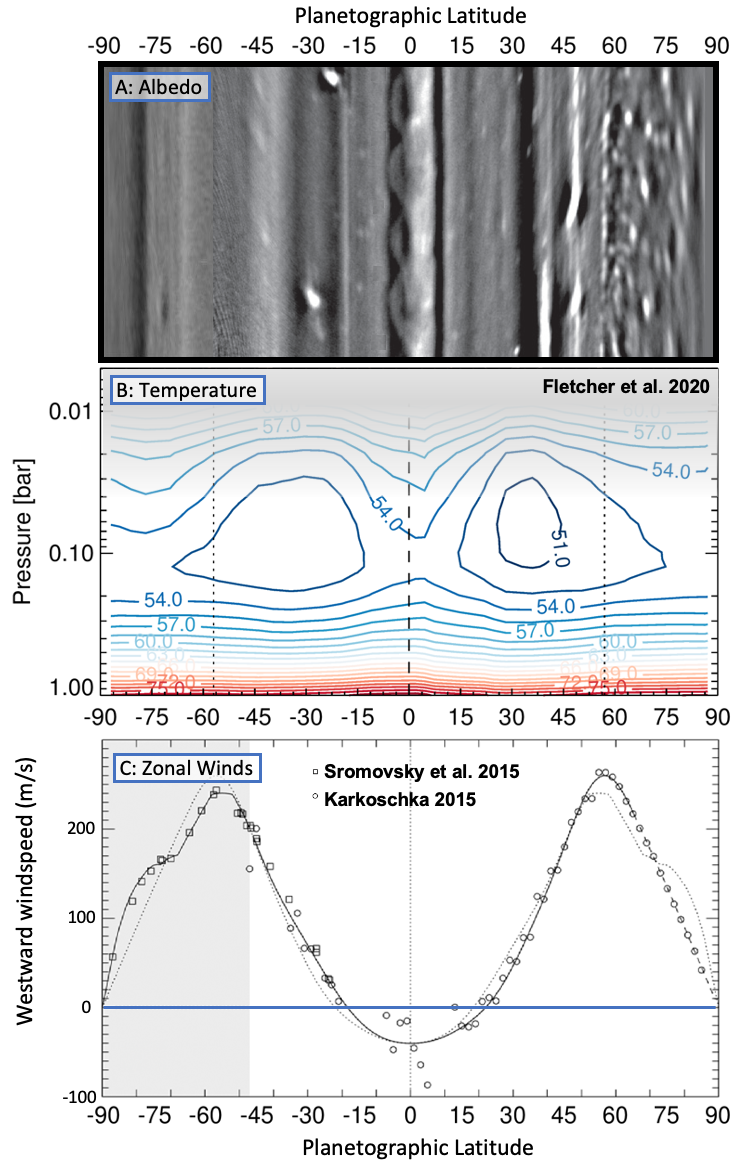}
\caption{Comparison of Uranus' near-IR reflectivity (A) to the zonal-mean tropospheric temperatures (B) and cloud-tracked winds (C).  In panel (A) we Voyager-2 imagery poleward of $60^\circ$S \citep{15karkoschka} with Keck H-band imagery in 2012 northward of $60^\circ$S \citep{15sromovsky}.  Panel (B) comes from inversion of Voyager/IRIS 25-50$\mu$m spectroscopy \citep{20fletcher_icegiant}, the grey shaded regions represents the declining information content at low pressures.  Zonal winds in (C) are adapted from Fig. 15 of \citet{15sromovsky}, combining Voyager-2 observations in the grey shaded area \citep[squares,][]{15karkoschka} with Keck observations (circles) at other latitudes, and the dotted curve represents a profile if reflected at the equator.} 
\label{temp-winds}
\end{center}
\end{figure}

Tropospheric zonal winds are constrained by tracking of discrete cloud features by Voyager 2 \citep{86smith, 15karkoschka}, the Hubble Space Telescope \citep{98karkoschka, 01hammel}, and ground-based facilities like Keck and Gemini \citep{05sromovsky, 05hammel, 09sromovsky, 15sromovsky}.  These sources were used to develop a `canonical wind profile' by fitting Legendre polynomials to the measured drift of features \citep[Fig. \ref{temp-winds}c,][]{15sromovsky}, although there remains some dispersion and potential temporal variations around these measurements.  Most of the tracked cloud features on Uranus are near the 1.3-bar methane condensation level or in the deeper 2-4 bar main clouds of H$_2$S ice \citep{18irwin_h2s}, but some can reach the 250-600 mbar region in the upper troposphere \citep{07sromovsky, 12sromovsky, 18roman}.  Some of the smaller-scale clouds evolve quite rapidly \citep{17irwin}, making it difficult to track them over multiple rotations.  Nevertheless, Fig. \ref{temp-winds} shows that the cloud tracking has revealed a single, broad retrograde jet at Uranus' equator, and a single high-latitude $\sim260$ m/s prograde jet in each hemisphere near $50-60^\circ$, quite unlike the multi-jet circulations of Jupiter and Saturn \citep[see the  reviews by][]{18sanchez_jets}.   Note that it is common practice to define Uranus' westward winds as positive (prograde), and the eastward equatorial jet as negative (retrograde), unlike on the other giants.  Furthermore, reconstruction of the gravity field measured by Voyager 2 \citep{91hubbard} out to the fourth order harmonic \citep{13kaspi} suggests that these tropospheric winds are confined to a layer approximately 1000 km deep (i.e., down to $~2$ kbar).

Geostrophic balance suggests that we should measure a correlation between horizontal temperature gradients and the vertical shear on the zonal winds \citep{04holton}.  The thermal banding observed in the microwave in Fig. \ref{ura_montage} \citep{20molter} is too fine-scale to be related to the broad structure of the zonal flow \citep{15sromovsky}, but thermal-IR observations by Voyager 2 \citep{87flasar, 98conrath} and ground-based facilities \citep[Fig. \ref{ura_montage},][]{15orton, 20roman} reveal larger-scale structure, with cool mid-latitudes in the 80-800 mbar range, contrasted with a warmer equator and poles.  The tropospheric temperature patterns in Fig. \ref{temp-winds}b appear to be unchanging in the time since Voyager 2, and imply maximum positive windshears near $\pm(15-30)^\circ$ latitude on the flanks of the equatorial retrograde jet, and maximum negative windshear near the high-latitude prograde jets at $\pm(60-75)^\circ$ \citep{20fletcher_icegiant}.  The windshear is minimal (i.e., close to barotropic, with no wind variability with height) in the $\pm30-50^\circ$ latitude range associated with the mid-latitude temperature minima, which is also where the most notable storm activity occurs (Section \ref{meteorology}).  Taken together, this suggests that the cloud-top prograde jets of \citet{15sromovsky} should decay with altitude, potentially due to frictional damping related to the breaking of vertically propagating waves in the upper troposphere \citep{87flasar}.  The equatorial winds would be more complicated, with the most rapid decay of the retrograde flow with height occurring away from the equator.  Furthermore, existing measurements of latitudinal temperature gradients and zonal winds have relatively poor spatial resolution, so it is reasonable to assume that finer-scale temperature/wind bands exist throughout the troposphere, as suggested by the reflectivity (Fig. \ref{temp-winds}a) and microwave emission maps.  Future missions with multi-spectral mapping capabilities are needed to resolve this.

Finally, there are hints in Fig. \ref{temp-winds} that both the zonal winds \citep{15sromovsky} and the tropospheric temperatures \citep{98conrath, 15orton} exhibit a north-south asymmetry, and that this asymmetry might be weakening with time.  Southern summertime winds at $50-90^\circ$S \citep{15karkoschka} appear to be very different to corresponding northern latitudes $60-83^\circ$N during spring \citep{15sromovsky}, where zonal drift rates adhere to solid-body rotation.  Long-term tracking of the winds will be required to understand whether this asymmetry is seasonal, or a permanent feature.

\subsubsection{Composition Contrasts}
\label{composition}

\begin{figure}
\begin{center}
\includegraphics[angle=0,width=0.5\textwidth]{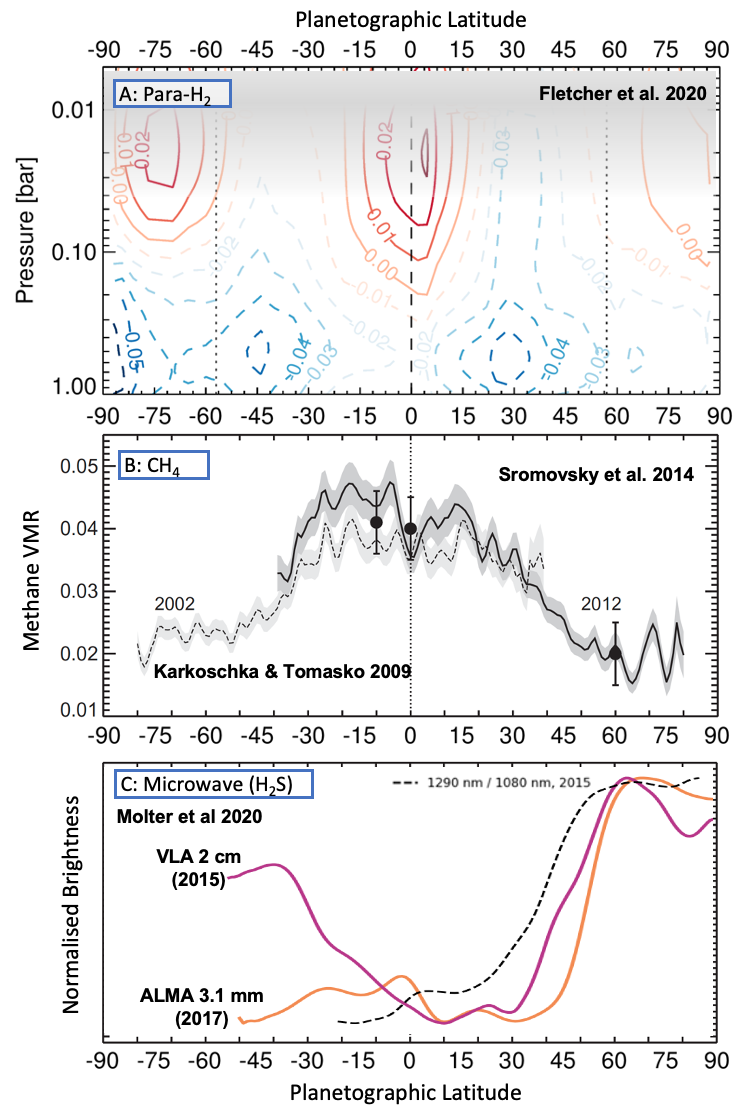}
\caption{Zonal-mean composition gradients in para-H$_2$ disequilibrium (A), tropospheric CH$_4$ (B), and tropospheric microwave emission indicating H$_2$S variations (C).  Panel (A) shows the extent of disequilibrium for para-H$_2$ (sub-equilibrium in dotted lines indicating upwelling, super-equilibrium in solid lines indicating subsidence) derived from Voyager/IRIS spectra \citep{20fletcher_icegiant}. Grey shading indicates the lack of information content for $p<70$ mbar.  The latitudinal distribution of methane in panel (B) is from \citet{09karkoschka, 14sromovsky}.  The deep distribution of H$_2$S is represented by the normalised microwave brightness in panel (C), as measured by ALMA and the VLA \citep{20molter}, where higher emission indicated depleted H$_2$S abundances, and the dashed line is indicative of the poleward CH$_4$ depletion as given in panel (B).} 
\label{ura_comp}
\end{center}
\end{figure}

Besides temperatures, winds, and aerosols, we also observe planetary banding in the gaseous composition in Fig. \ref{ura_comp}.  Fine-scale banding in microwave emission in Fig. \ref{ura_comp}c \citep{20molter} has already been discussed, and can be interpreted in terms of variations in H$_2$S abundance as a function of latitude.  The collision-induced continuum measured by Voyager 2 also reveals the latitudinal contrasts in para-hydrogen, the even spin isomer of H$_2$ \citep{87flasar, 98conrath, 20fletcher_icegiant}.  The para-H$_2$ fraction is quenched at 25\% in the warmer deep troposphere, and chemical equilibration is sufficiently slow that upward motion can bring this low-para-H$_2$ air upwards into the cooler upper troposphere, where the equilibrium para-H$_2$ abundance should be much larger.  Sub-equilibrium para-H$_2$ in Fig. \ref{ura_comp}a is therefore indicative of upwelling, which is seen at mid-latitudes in the 80-800 mbar region.  Conversely, super-equilibrium para-H$_2$ is observed at the equator and poles, which is taken to be an indicator of subsidence from the cold tropopause.  Caution is required though, as poorly-understood catalysis processes on aerosol surfaces could cause the equilibration timescale to vary with location \citep{82massie}. 

Small-scale contrasts in Fig. \ref{ura_comp}b-c are superimposed onto a significant equator-to-pole decline in the volatile abundances CH$_4$ and H$_2$S.  Uranus' south polar depletion of methane was first detected via Hubble 300-1000 nm spectroscopy in 2002 \citep{09karkoschka}, when the north polar region was still largely hidden from view.  After equinox in 2007, further Hubble observations in 2012 \citep{14sromovsky} and 2015 \citep{19sromovsky} revealed that the north polar region exhibited a similar depletion, and that the equator-to-pole gradient from $\sim4$\% near the equator to $\sim2$\% at high latitudes appeared to be relatively stable over time.  Furthermore, this gradient exhibited a step-like structure, with a transition in abundance near the $\pm30-40^\circ$ latitude region, although degeneracies with the assumed aerosol distribution can modify the derived gradients.  Uranus' polar methane depletion was also observed independently from ground-based facilities in the near-infrared, including the IRTF \citep{13tice}, Keck \citep{15dekleer}, Palomar \citep{18roman}, the VLT and Gemini \citep{18toledo, 18irwin_h2s}.  \citet{19sromovsky} reproduced the spectra with a methane distribution that was latitudinally uniform for $p>5$ bar (in the 2.7-3.5\% range, depending on aerosol scattering properties), and only latitudinally variable at lower pressures in the upper troposphere, with a factor of three decrease in the upper tropospheric CH$_4$ from $30^\circ$N to $70^\circ$N.

However, the implied equator-to-pole circulation, with rising methane-rich air at low latitudes, moving polewards as CH$_4$ is removed by condensation, precipitation, and sedimentation, before methane-poor air sinks at high latitudes, may in fact extend much deeper.  VLA centimetre observations of Uranus between 1982 and 1994 \citep{88depater, 91depater_radio, 03hofstadter}, probing down to 50 bar, revealed Uranus' microwave-bright south polar region due to a strong depletion absorbers, with a strong brightness gradient near $45^\circ$S (Fig. \ref{ura_montage}).  VLA observations since 2003 show that the north polar region was similarly bright \citep{04hofstadter_dps, 18depater, 20molter}, and this high brightness can be primarily explained by the absence of H$_2$S down to $\sim35$ bars \citep{20molter}.  Whether this polar depletion of deep H$_2$S is caused by the same circulation and subsidence responsible for the shallow CH$_4$ depletion is an open question \citep{14sromovsky, 20fletcher_icegiant}.  Furthermore, neither the methane nor the H$_2$S appear to be tracing the upper tropospheric circulation revealed by the para-H$_2$ and temperature distribution, which would suggest mid-latitude upwelling, rather than equatorial upwelling.   Indeed, polar subsidence would tend to inhibit convection and CH$_4$ cloud formation \citep{12sromovsky}, and yet discrete clouds were clearly visible at Uranus' north pole during early northern spring in Fig. \ref{ura_voyager} \citep{14sromovsky}.

\subsubsection{Meridional Overturning}

Uranus' tropospheric circulation has yet to be explored via detailed numerical simulation, but the differences between (i) the broad contrasts in upper tropospheric temperatures, winds and para-H$_2$; (ii) the fine-scale banding observed in the aerosols and deep thermal emission; and (iii) the strong equator-to-pole contrasts in CH$_4$ and H$_2$S, imply that circulation cells might be restricted to two or more discrete tropospheric layers \citep{15sromovsky, 20fletcher_icegiant}.  Similar `stacked circulation cells' have been proposed for Jupiter and Saturn \citep{00ingersoll, 05showman, 20fletcher_beltzone}, and it is possible that the stabilising influence of molecular weight gradients where clouds condense might help in separating these layers, creating stable regions where convection is inhibited \citep{95guillot, 17leconte, 17friedson}, and possibly preventing vertically-extended convection and circulation entirely.  Indeed, thin-layered convection was proposed to explain Voyager observations of Uranus' lapse rate and para-H$_2$ distribution by \citep{87gierasch}.  The CH$_4$ and H$_2$S condensation levels may be good candidates for transitional layers,  decoupling motions above and below \citep{19hueso}.  Such differences in atmospheric circulation as a function of depth remain to be proven, but the transition from a deep domain where eddies provide momentum to the zonal jets, to a shallow domain where zonal jets decay with altitude due to unidentified frictional drag (i.e., a radiative- or wave-driven circulation regime), remains a plausible explanation for the multi-spectral observations available today.

Finally, we note that the tropospheric circulation may also have implications for Uranus' stratosphere.  Stratospheric chemistry will be discussed in Section \ref{chem}, but we note that photolysis of stratospheric methane produces acetylene, C$_2$H$_2$, which is the only stratospheric chemical to have been mapped on Uranus to date \citep{20roman}.  Using images at 13.7 $\mu$m, \citet{20roman} revealed excess C$_2$H$_2$ emission at mid-latitudes, counter to the expectations of seasonal photochemical models \citep{18moses}.  They proposed two equally plausible mechanisms:  either tropospheric mid-latitude upwelling was transporting methane-rich air through the cold trap and into the stratosphere, where it was subsequently photolysed to form C$_2$H$_2$; or the stratosphere exhibited subsidence and adiabatic warming at mid-latitudes.  The first case can be considered as an upward extension of the upper tropospheric circulation, whereas the second case is a circulation where the mid-latitude stratospheric subsidence opposes the mid-latitude tropospheric upwelling, yet another tier of a stacked circulation system.  Distinguishing between these two possibilities is a topic of ongoing research.

\subsection{Uranian Meteorology}
\label{meteorology}

\subsubsection{Convective Storms}
Discrete meteorological features, from small bright clouds, to planet-encircling waves, giant storms and dark vortices, are superimposed upon and intricately connected to the large-scale atmospheric circulation described in Section \ref{circulation}.  Some of the small-scale bright clouds, presumably comprised of methane ice reaching pressures of 300-600 mbar, high above the main H$_2$S cloud deck at 2-4 bars, may be convective in origin, driven by a combination of latent heat release as CH$_4$ condenses \citep{89stoker, 20hueso}, and potentially by the energy released during the interconversion between para- and ortho-H$_2$, which have different specific heats \citep{95smith}.  However, Uranus shares the same problem as the other giant planets, in that air saturated with condensates will be heavier than the surrounding `dry' H$_2$-He mixture \citep{95guillot, 17leconte, 17friedson}.  As described in Section \ref{circulation}, the strong vertical gradient in molecular weight prevents convection extending over tremendous heights, instead limiting it to vertically-thin layers \citep{87gierasch}.  Strong initial perturbations are therefore required to counteract this static stability (effectively a negative buoyancy) to drive self-sustaining CH$_4$-rich updrafts, but this has yet to be verified observationally \citep{19hueso, 20hueso}.  Indeed, \citet{19guillot} point out that this moist CH$_4$-driven convection is occurring at shallow pressures (0.1-1.5 bars) and low optical depths compared to the deep, hidden H$_2$O clouds of Jupiter and Saturn, meaning that Uranus (and Neptune) provide ideal destinations to investigate how convection operates on hydrogen-rich worlds where it is inhibited by the weight of the condensables.  Furthermore, Uranus' moist convection might be rather different to Jupiter's, as H$_2$O will play a very limited role in the observable upper troposphere, and H$_2$S condensation provides only a limited source of latent heat to drive convection.

\citet{20hueso} define Uranus' convective features to be storm features that show divergence above the clouds over short timescales, but they point out that the frequency and distribution of such storms is not well known, and much of the observed cloud activity might not be convective.  Instead, some bright features may be orographic structures associated with pressure perturbations at deeper levels (e.g., those associated with dark, hidden vortices), which do not require buoyancy to produce condensation, as on Neptune \citep{01stratman}.   A particularly long-lived feature at southern mid-latitudes, some 5000-10000 km wide, was known as the `Berg` and had bright features rising to the 550-750 mbar level, but with the main parts of the structure near 1.7-3.5 bars \citep{11depater}.  The berg was captured by HST and Keck between 1994 and 2009 \citep{09sromovsky, 11depater, 05hammel, 15depater, 15sromovsky}, and may even have been observed by Voyager in 1986 \citep{05sromovsky}.  It oscillated in latitude (around $35.2^\circ$S) and longitude, migrating towards the equator after 2005, reaching $27^\circ$S in 2007, and disintegrating when it reached $5^\circ$S in late 2009 \citep{09sromovsky, 11depater}.  The Berg exhibited intense brightenings in 2004 and 2007 suggestive of convective storms within the feature \citep{15depater}, but there was no evidence that it was related to a deeper unobserved vortex \citep{19hueso}.

Discussion of strong convection naturally raises questions about the potential for lightning.  Methane is non-polar, but H$_2$S, NH$_3$, NH$_4$SH and H$_2$O provide the mixed-phase materials that could produce lightning \citep{20aplin}.  Microphysical modelling suggesting that the radio emissions discovered by Voyager \citep{86zarka} are more likely related to the NH$_4$SH cloud than the deep water layers \citep{20aplin}.

\subsubsection{Discrete Cloud Activity}

Uranus has exhibited a range of discrete cloud phenomena since the early Voyager and Hubble observations.  Eight features were observed between $35-70^\circ$S by Voyager \citep{86smith} near northern winter solstice.  As winter proceeded in the mid-1990s, notable bright cloud features were observed at mid northern latitudes \citep{98karkoschka}, as they were emerging into sunlight.  Since 2000, ground-based facilities like Keck were frequently detecting cloud features \citep{01hammel, 02depater, 05hammel}, including some in 2004 that reached sufficiently high altitude to be detected in the strong CH$_4$ band near 2.12 $\mu$m \citep{05hammel_ura04}.  Northern mid-latitudes $28-42^\circ$N appear to be intrinsically active \citep{07sromovsky_uraclouds}, particularly in the years following northern spring equinox ($L_s=0^\circ$, 2007), with clouds reaching 300-500 mbar, and new outbreaks observed in 2004-06 and 2011 \citep{07sromovsky, 09sromovsky, 12sromovsky}.  A record-breaking bright storm occurred at $15^\circ$N in 2014, as shown in Fig. \ref{ura_montage} \citep{15depater}, thought to be formed from a complex of smaller storms.  And another at $32^\circ$N possessed a longitudinally-elongated tail in the 1-2 bar range, qualitatively similar to storm tails observed on Jupiter and Saturn \citep{15depater, 15sromovsky, 16irwin_ura, 17irwin}.  

It is natural to ask whether Uranian storms exhibit a temporal dependency, given that some of the brightest activity appeared to occur in the hemisphere emerging from the darkness of winter.  Alternatively, maybe Uranian storms occur episodically due to powerful outbursts followed by long periods of quiescence as CAPE (convective available potential energy) accumulates, as suggested for Saturn's annual storms \citep{15li}.  However, storm statistics are currently insufficient, and the observations are biased to the sunlit hemisphere, so robust assessment of the destabilising influence of increased insolation remains to be performed.  The one exception is in the polar domain in Fig. \ref{ura_voyager}, where the south pole displayed no discrete activity during southern summer \citep{86smith,15karkoschka}, whereas clusters of bright spots with 600-800 km diameters were observed in the north polar region after it emerged into sunlight in northern spring \citep{12sromovsky, 15sromovsky}.  This notable asymmetry suggests some form of convective inhibition (or obscuration by overlying hazes) developing as spring turns to summer, something which will be testable as Uranus approaches northern summer solstice ($L_s=90^\circ$) in 2030.

\subsubsection{Vortices and Waves}

The vast majority of discrete features tracked on Uranus have been bright cloud features, rising above the main cloud decks.  However, Uranus also infrequently exhibits dark anticyclonic ovals, which can form and dissipate on the timecales of years.  Hubble was the first to observe a dark vortex in 2006 at $28^\circ$N \citep{09hammel}, and since that time several more have been observed \citep[e.g.,][]{15sromovsky}.  These are typically smaller than those observed on Neptune, and whether they are accompanied by bright (and more readily visible) companion clouds due to air being forced up and over the vortex \citep[orographic clouds,][]{01stratman} remains a topic of ongoing exploration.  It is not clear whether the dearth of anticyclonic ovals on Uranus with respect to Neptune is an observational bias (e.g., due to different overlying aerosols and gas absorption), or a real difference between the two worlds.

Finally, Uranus exhibits wave phenomena that manifest as longitudinal reflectivity contrasts in the troposphere.  A chain of diffuse bright features just north of the equator, first suggested by Keck images in 2003 \citep{05hammel}, were observed by Keck in 2012 \citep{15sromovsky} to be spaced every $30-40^\circ$ longitude.  A second wave was captured by the 2012-14 Keck observations using adaptive optics and derotation techniques \citep[Fig. \ref{temp-winds}a,][]{12fry, 15sromovsky}, revealing a transverse `scalloped' equatorial wave pattern, just south of the equator, with diffuse bright features every $19-21^\circ$ longitude (wavenumber 17-19) and a latitudinal amplitude of $2.4-2.9^\circ$.  \citet{15sromovsky} suggest that Kelvin waves or mixed internal gravity-Rossby waves may be at work, but that this cannot be properly characterised until the dispersion relation (phase speed versus wavelength) is determined. 

Wave phenomena are unlikely to be restricted to Uranus' troposphere, and may also be present in the stratosphere, modulating emission from stratospheric hydrocarbons \citep{20roman} and generating rotational variability in disc-averaged mid-IR spectra observed by Spitzer \citep{21rowegurney}.

\subsection{Chemistry}
\label{chem}

Although the thermochemistry and condensation chemistry of Uranus's troposphere were described in Section \ref{partone}, the influence of UV photolysis, coupled with the atmospheric circulation (Section \ref{circulation}), can contribute to the sources, sinks, and spatio-temporal distributions of chemicals detected in Uranus' upper troposphere and stratosphere.  Whilst much of this section will be devoted to stratospheric chemistry, photolysis of the yet-to-be detected PH$_3$ \citep{09moreno_dps, 19teanby} and H$_2$S lofted to high altitude may contribute to upper tropospheric hazes, and potentially more so than on Jupiter and Saturn because of the absence of NH$_3$ to act as a UV shield \citep{05visscher, 20moses}.        

\subsubsection{Methane Photochemistry}

\begin{figure*}
\begin{center}
\includegraphics[angle=0,width=1.0\textwidth]{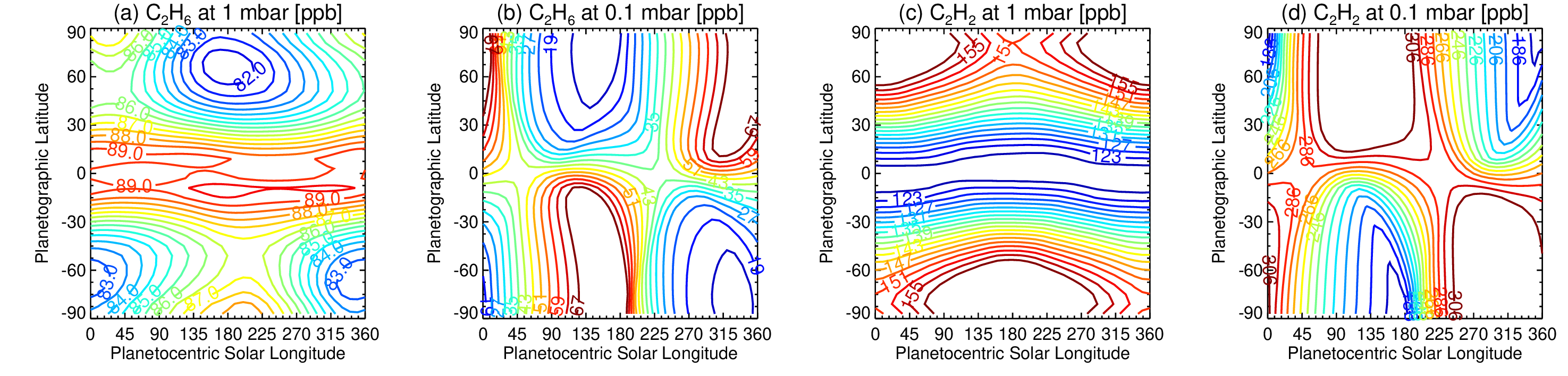}
\caption{Predicted contrasts in abundances of stratospheric hydrocarbons ethane (C$_2$H$_6$) and acetylene (C$_2$H$_2$) at 0.1 and 1 mbar in parts per billion.  Each panel shows the abundance as a function of latitude and planetocentric solar longitude, revealing larger variations at lower pressures, and the different behaviours of these two species.  Plotted from the seasonal photochemistry model of \citet{18moses}.}
\label{cxhy}
\end{center}
\end{figure*}

Methane photolysis products dominate the stratospheric composition of both Ice Giants \citep{91atreya}, but the strength of vertical mixing by eddy diffusion (weak on Uranus, strong on Neptune) has stark consequences for the distribution of species, with Uranian methane photochemistry operating in a higher-pressure regime than on any other giant planet \citep{18moses}.  The lower CH$_4$ homopause on Uranus \citep{87herbert} has the surprising consequence that seasonal contrasts are more muted than on Neptune, and that CH$_4$ chemistry is not as well coupled to exogenic oxygenated materials deposited at lower pressures (see Section \ref{oxygen}).  Nevertheless, photochemistry on Uranus results in a complicated mix of hydrocarbons \citep{83atreya, 05moses, 10dobrijevic, 18moses} that can be investigated via mid-infrared remote sensing \citep{87orton, 97feuchtgruber, 98encrenaz, 14orton, 20roman, 21rowegurney} and UV occultations \citep{87herbert, 90bishop}.  Acetylene (C$_2$H$_2$) was first discovered by ISO \citep{98encrenaz}; ethane (C$_2$H$_6$), methyl-acetylene (C$_3$H$_4$) and diacetylene (C$_4$H$_2$) were observed by Spitzer \citep{06burgdorf, 14orton_chem}.  Some products, such as ethylene, propane, benzene and methyl have not yet been detected \citep{20moses}.  The effects of the high-pressure photochemistry are apparent in mid-IR observations:  Uranus' hydrocarbons are confined to altitudes below the $\sim$0.1-mbar level in Fig. \ref{temp-clouds}, and the ratio of ethane to acetylene is very different on Uranus compared to all the other giants \citep{14orton_chem}.

The latitudinal distribution of these species is expected to vary due to stratospheric circulation and seasonal photochemistry (Fig. \ref{cxhy}).  Although photochemistry occurs primarily in UV sunlight (145 nm), and thus photolytic rates will be largest at polar latitudes that receive a higher annual-average solar insolation than the equator, small amounts of solar Lyman alpha scattered from hydrogen in the local ISM can provide a secondary photolysis source during the darkness of winter.  Chemical abundances respond quickly at low pressures to changes in the UV flux, but Fig. \ref{cxhy} shows that contrasts should decrease at higher pressures, where diffusion and chemical timescales increase, producing subtle hemispheric asymmetries at $p<1$ mbar with phase lags in response to insolation changes \citep{18moses}.  Thus short-lived species like acetylene, methyl-acetylene and diacetylene are expected to have maximum abundances at the summer/autumn poles, whereas long-lived species like ethane are expected to peak near the equator because photolytic destruction effectively competes with production in the high-latitude summer.  However, these predictions assume uniform distributions of stratospheric CH$_4$ as the source material, which may not be the case given the strong equator-to-pole CH$_4$ gradients in the troposphere (Section \ref{composition}).  Furthermore, there remains some disagreement about what the stratospheric CH$_4$ abundance actually is, with a factor of six in stratospheric abundance between Spitzer results in 2007 \citep{14orton_chem} and Herschel results in 2011 \citep{15lellouch}.  Both inferences were from disc-integrated spectra, so this could be a sign of changing viewing geometry, latitudinal temperature gradients, changes in efficiency of vertical mixing, temporal variability, or some combination of the above.  

Of these species, only emission from C$_2$H$_2$ has been mapped across Uranus, showing stark discrepancies from the photochemical model predictions \citep{18moses, 20roman}, and implying a key role for circulation: either strong mid-latitude upwelling of CH$_4$-rich air from the troposphere; or mid-latitude subsidence in the stratosphere (Section \ref{circulation}).  The ground-based VLT maps of stratospheric C$_2$H$_2$ emission in Fig. \ref{ura_montage} suggest a distinct equatorial minimum at latitudes $<\pm25^\circ$ \citep{20roman}, counter to photochemical expectations and the complete opposite of what is observed in the tropospheric temperatures.

The products of CH$_4$ photochemistry, such as benzene, acetylene, ethane and propane (in addition to exogenic H$_2$O and CO$_2$) can condense to form thin haze layers in the 0.1-30 mbar range (Fig. \ref{temp-clouds} and \ref{clouds}), some of which are visible in high-phase imaging from Voyager \citep{91rages, 93romani, 17moses, 18toledo} and may be modulated by vertically propagating waves.  The condensed hydrocarbons could also sediment downwards to coat tropospheric hazes, or serve as cloud-condensation nuclei for CH$_4$ and H$_2$S condensation at higher pressures.

\subsubsection{Coupling to External Oxygen}
\label{oxygen}

Oxygen species like CO, CO$_2$ and H$_2$O are present in the upper stratospheres of each of the giant planets \citep{97feuchtgruber, 17moses}, originating from cometary impacts, satellite debris, and ablation of interplanetary dust and ring particles.  What makes Uranian photochemistry particularly interesting is that this oxygen deposition is separated from the methane photochemistry due to the sluggish vertical mixing, making it an intriguing counterpoint to the other giants \citep{17moses}.  Uranus' stratospheric water was detected by ISO \citep{97feuchtgruber}; CO from the fluorescent emission in the infrared \citep{04encrenaz} and sub-millimeter emission \citep{13cavalie_ura}; and CO2 from Spitzer \citep{06burgdorf, 14orton}.  Photolysis of CO and CO$_2$ can lead to secondary peaks of hydrocarbon production at high altitudes above the levels where the CH$_4$ abundance is dropping away \citep{18moses}, whereas H$_2$O will condense to form a 10-mbar ice haze. An interplanetary dust particle source is sufficient to explain the observed amount of CO, CO$_2$ and H$_2$O in Uranus' stratosphere \citep{17moses}, although cometary impacts could also contribute \citep{19lara}.  

\subsubsection{Radiative Balance}

The vertical distribution of hydrocarbons, particularly CH$_4$, C$_2$H$_2$ and C$_2$H$_6$, determine the radiative heating and cooling in Uranus' middle atmosphere.  For this reason, the influence of thin hazes and secondary peaks of hydrocarbon production can have a significant influence.  Uranus' upper troposphere and stratosphere is warmed by shortwave absorption by CH$_4$ and aerosols, and cooled by longwave emission from C$_2$H$_2$ and C$_2$H$_6$ in the stratosphere, and from the collision-induced H$_2$-He continuum in the troposphere \citep{98conrath, 18li}.  However, it has long been noted that absorption of weak sunlight by CH$_4$ alone cannot balance the efficient cooling from H$_2$ and, to a lesser extent, the hydrocarbons, so the middle atmosphere is warmer \citep{14orton} than expected.  Additional sources of heat from vertically propagating gravity waves or the radiative contribution of hazes have been invoked to explain the discrepancy between observations and predictions \citep{18li}.

Uranus' extreme obliquity, resulting in a larger irradiance at the poles than the equator, could also yield seasonal temperature asymmetries.  However, Uranus' radiative time constant in the troposphere is longer than the Uranian year \citep{87friedson, 90conrath}, such that tropospheric temperatures are not expected to change significantly with time, and to track the annual-average equilibrium values \citep{93lunine}.  This is consistent with the lack of observed tropospheric temperature variation between Voyager (Fig. \ref{temp-winds}) and the present day \citep{15orton, 20roman}.  Stratospheric temperature asymmetries have never been measured because of the extreme cold and resulting low mid-IR radiance, but we note that Uranus' stratospheric radiative time constant is actually longer than that on Neptune, due to the low CH$_4$ abundance and cold temperatures, such that stratospheric temperature asymmetries might be weaker on Uranus and again simply track the annual average (i.e., warmer at the poles, and cooler at the equator).  Future missions capable of measuring temperature contrasts between the summer and winter hemisphere are needed to address this question.

\subsection{Temporal Change}
\label{temporal}

\begin{figure*}
\begin{center}
\includegraphics[angle=0,width=1.0\textwidth]{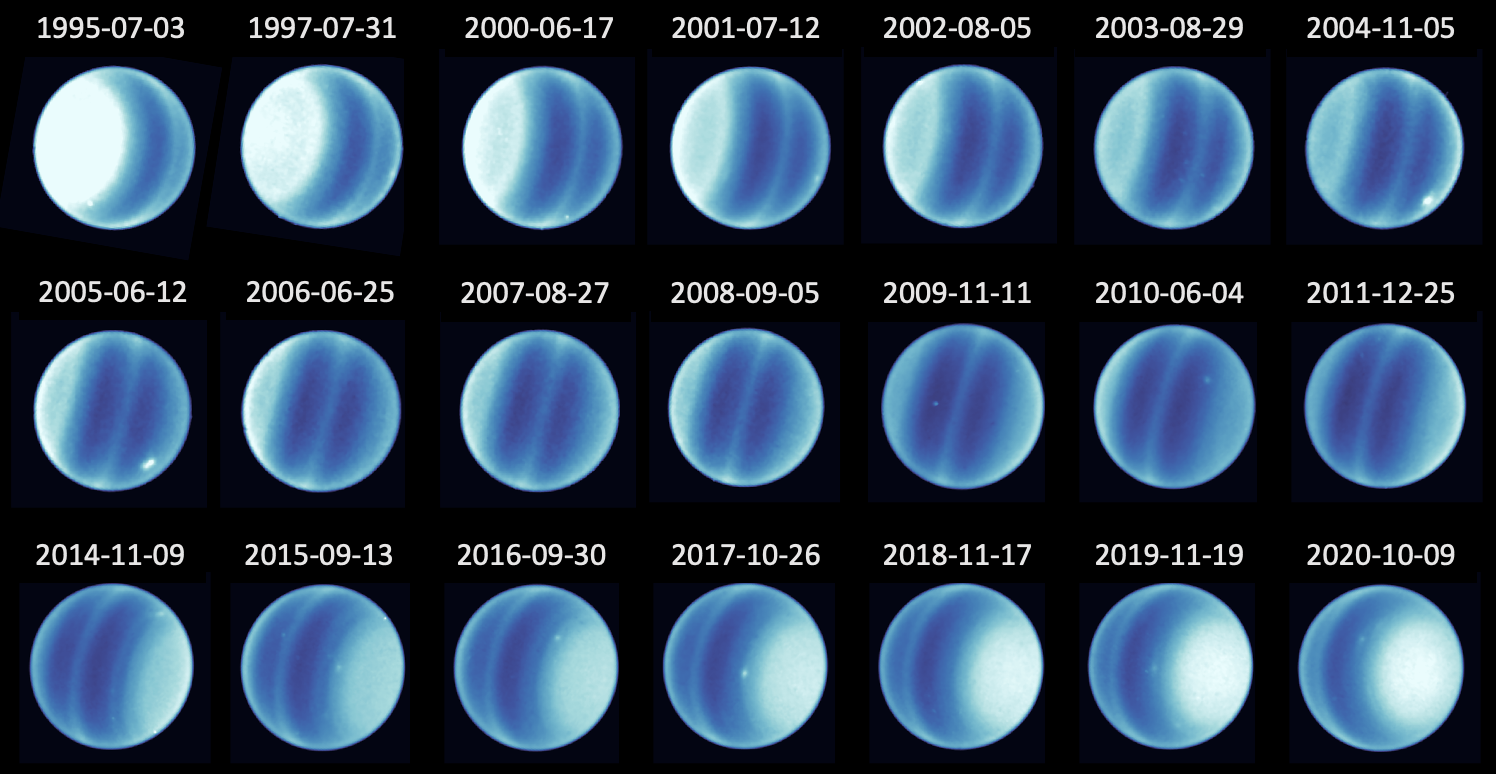}
\caption{Hubble observations at 619 nm, a weak methane absorption band, showing the evolution of Uranus' reflectivity over a quarter of a century, spanning mid-winter, spring equinox, and mid-spring in the northern hemisphere.  Observations before 2009 were acquired by the WFPC2 instrument, observations after 2009 were acquired by the WFC3 instrument, largely by the Outer Planet Atmospheres Legacy (OPAL) programme since 2014. }
\label{hubble}
\end{center}
\end{figure*}

A theme running through our discussion of Uranus' circulation (Section \ref{circulation}), meteorology (Section \ref{meteorology}) and chemistry (Section \ref{chem}), is that we can learn more about a system by observing how it evolves with time.  Given that resolved observations of Uranus' atmosphere spans almost four decades, and photometric observations span even longer, Uranus is the ideal testbed for studying variability on an Ice Giant.  We have already described three time-variable aspects of Uranus' atmosphere.  Firstly, predictions of radiative-climate models and diffusive photochemistry models suggest that upper tropospheric and stratospheric temperatures and hydrocarbon distributions might show subtle hemispheric asymmetries that reverse as seasons progress, but that the relevant time constants are so long that such contrasts are expected to be small.  Nevertheless, subtle mid-latitude asymmetries in tropospheric temperatures \citep{98conrath} and winds \citep{15sromovsky} have been observed.   Secondly, high-latitude winds were found to be different between northern spring in 2012-14 \citep[where a broad region of solid-body rotation from 62-83$^\circ$N was identified,][]{15sromovsky} and southern summer in 1986 \citep[with a much smaller region of solid-body rotation,][]{15karkoschka}.  Whether this is a permanent asymmetry, or something that will change as northern summer solstice approaches in 2030, remains to be seen.  Thirdly, there is evidence that small-scale cloud and storm activity varies over time, potentially as a result of the changing atmospheric stability as the seasons change, and this is particularly notable in the polar domain \citep{12sromovsky}.

These three time-variable processes are dwarfed by long-term variations associated with Uranus' seasonal polar cap and collars (Fig. \ref{hubble}), which are likely related to the processes discussed above and which modulate the disc-integrated photometry as seen from Earth \citep{19lockwood}.  Voyager observations during southern summer revealed a reflective polar cap poleward of $45^\circ$S, originally thought to be due to an increase in the optical depth of the CH$_4$ clouds near 1.2-1.3 bars \citep{91rages}. In the years after Voyager, Hubble observations between 1994 and 2002 revealed a darkening of the south pole (Fig. \ref{hubble}), the formation of a bright ring near $70^\circ$S and a south polar collar at $45^\circ$S \citep{04rages}.  As the northern hemisphere came into view after 2007, the south polar collar diminished \citep{10irwin, 18roman}, and a north polar collar developed at $45^\circ$N \citep{12irwin}, with a bright `north polar cap' observed after 2014 \citep{15depater, 15sromovsky, 18toledo}.  By computing photometric brightness from Hubble imaging between August 1994 and October 2015, \citet{16karkoschka_dps} showed that both the changing view of Uranus in Fig. \ref{hubble} and the real physical changes were modulating the brightness - darkening of high southern latitudes, and brightening of high northern latitudes. 

What could be causing these changes at high latitudes?  CH$_4$ is strongly depleted at the poles so there is less absorption (Section \ref{composition}), and although \citet{18toledo} suggested that temporal evolution of this `methane hole' could be responsible, \citet{19sromovsky} revealed that the polar CH$_4$ was relatively stable over time, and suggest that increased scattering from seasonally-changing aerosols is required.  There are problems with this, as the polar aerosols appear to be changing over short timescales in Fig. \ref{hubble}, so cannot be explained via long radiative \citep{90conrath}, photochemical \citep{18moses}, or microphysical processes \citep[aerosol accumulation and sedimentation are slow processes that would produce a substantial seasonal lag,][]{19toledo_ura}.  It seems likely that tropospheric and stratospheric circulation must be playing a role \citep{20fletcher_icegiant}, again highlighting the complexity of the coupled circulation, chemistry, and clouds of Uranus' atmosphere.  

Long-term monitoring of visible albedo (472 and 551 nm) between 1972 and 2016 \citep{99lockwood, 07hammel, 19lockwood}, and broadband B and V photometry from 1950 to 1966 \citep{06lockwood}, are consistent with the geometric effects as the bright polar caps and collars come in and out of view, but also suggest more subtle, secular variations in brightness.  In particular, Uranus' reflectivity is related to the 11-year solar cycle, both through changing levels of UV \citep[which could be modulating the colours of aerosols via a `tanning' process,][]{90baines}, and also through the ion-induced nucleation generated by galactic cosmic rays \citep{17aplin, 20aplin}.  These studies highlight the value of long-term Earth-based observations, to provide temporal context to short-lived spacecraft missions.

\subsection{Connections}
\label{connections}

\subsubsection{Connection to the Interior}

Although this review is focused on Uranus' atmosphere, it cannot be considered in isolation from the deeper interior, and the external charged particle environment.  It is not immediately clear where an Ice Giant atmosphere (molecular envelope) ends, and its interior or icy mantle begins.  Differential rotation due to zonal winds appears to be restricted to the outermost 1000 km of Uranus' 25559 km equatorial radius \citep[i.e., the outer 4\% down to $\sim2$ kbar,][]{13kaspi}.  Deeper down at $\sim20$ kbar and 1200 K, is it possible that H$_2$O is insoluble in H$_2$ \citep{13bali}, leading to the formation of a liquid water ocean at tens of kilobars \citep{15bailey_agu}.  This immiscibility can lead to sharp interfaces, or an ocean surface, within the interior \citep{21bailey}.  This ocean would transition from being non-conducting to ionic/superionic at hundreds of kilobars, and could be responsible for removing NH$_3$ by dissolution to explain the low N/S ratio detected in the atmosphere (Section \ref{partone}). The generation of Uranus' internal dynamo \citep{86ness} may be associated with convection in the partially dissociated fluid water layers \citep{11redmer, 13soderlund}, and appreciable conductivity can be achieved at 20-30\% of the radii below the surface \citep{20soderlund}.  Dynamo simulations predict large-scale circulation in this watery layer, with equatorial upwelling and, depending on the thickness of the convecting layer, polar circulation cells \citep{13soderlund}, although it is unclear how this might relate to the circulation in the molecular envelope in Section \ref{circulation}.  At even greater depths, high-pressure laboratory experiments suggest that water could form a thick superionic `icy mantle', creating a solid-phase viscous lattice of oxygen ions surrounded by a sea of free hydrogen \citep{13wilson, 18millot, 19millot}.  

These substantial transitions between exotic phases of matter in Uranus' interior could lead to inhibition of convection and retention of interior heat \citep[e.g.,][]{95hubbard}, significantly influencing the thermal evolution of Uranus \citep{11fortney}.  Indeed, different rates of hydrogen-water demixing, due to H$_2$O immiscibility and the formation of a water ocean, could explain the very different heat flow on Uranus compared to Neptune \citep{21bailey}.  Furthermore, the presence of these transitions could successfully decouple motions in the observable `dry' atmosphere from those in the deeper, water-rich interior oceans.  However, \citet{20helled_icegiant} provide a cautionary note that we still do not know the bulk oxygen content of Uranus, to the extent that it might instead be dominated by rocky materials rather than water, and implying that the oft-used term `Ice Giant' may be misleading.

\subsubsection{Connection to the Exterior}

The circulation and chemistry of the middle atmosphere (e.g., the stratosphere) will also be intricately connected to processes shaping the upper atmosphere (e.g., ionosphere/thermosphere), above the homopause where hydrogen dominates.  Weak solar heating alone cannot explain the high temperatures of Uranus' thermosphere \citep{87herbert}, a deficit known as the `energy crisis,' and contributions from auroral heating and potentially the breaking of vertically-propagating waves have been invoked to close the gap between observations and expectations.  Extreme UV radiation, or impact ionisation from the aurora, produces the H$_3^+$ ion, which was first detected on Uranus in 1992 \citep{93trafton}, and has been monitored ever since, revealing a long-term cooling of Uranus' thermosphere over three decades \citep{19melin}. That H$_3^+$ production appears to be efficient on Uranus, but not Neptune, may be related to the weak atmospheric mixing and low homopause height.  The thermospheric cooling trend appears to be decoupled from Uranus' geometric season (i.e., it did not change after the 2007 equinox), but might be explained by a `magnetic season:' differing amounts of Joule heating at different points in the planet's orbit, or changes to the homopause height with time, or a hemispherically asymmetric homopause that changes the H$_3^+$ density on the hemisphere visible from Earth \citep[see][for a review]{20melin}.  Whether this redistribution of energy in Uranus' upper atmosphere has any implications for the energy balance, circulation, and chemistry of the stratosphere remains to be seen.  

Auroral emission has been observed via excitation of H Lyman-$\alpha$ and H$_2$ bands due to precipitating energetic particles \citep{20lamy}.  These are located at magnetic poles tilted from the rotational axis by $\sim60^\circ$, as observed by Voyager UVS \citep{87herbert} and later Hubble \citep{12lamy, 17lamy}.  Auroral brightening has also been tentatively detected in the infrared via H$_3^+$ emission \citep{20thomas_epsc}.  On Jupiter ion-neutral chemistry related to the aurora can lead to a unique balance of stratospheric chemicals and aerosols in the polar domain, so future observations could test whether this is also true on Uranus, given the mid-latitude location of the magnetic poles.

\section{Conclusions and Outstanding Questions}
\label{conclude}
% Conclusion (400 – 500 words)
%  Draw together significant conclusions that assess the field, including strengths and weaknesses.
%  Conclude with your judgment on what significant questions remain, are being pursued, or should be pursued.

This review should dispel any remaining myths about Uranus' atmosphere held over from the first Voyager images, revealing it to be an extreme laboratory for testing our understanding of the processes shaping planetary atmospheres.  Indeed, Uranus' atmosphere occupies a unique regime in parameter space: 
\begin{itemize}
    \item An intermediate-sized world rotating a rate slower than Jupiter but faster than Earth, with implications for the banded structure (Section \ref{circulation});
    \item A hydrogen-rich molecular envelope with a bulk that is substantially enriched in heavy materials (volatiles and rocks) and strong equator-to-pole gradients in condensables (Sections \ref{partone} and \ref{circulation};
    \item The unusual seasonal effects due to both the extreme axial tilt, weak sunlight, and negligible internal heat, perhaps sequestered by strong layering associated with phase transitions in watery oceans and icy mantles at tens and hundreds of kilobars (Sections \ref{temporal} and \ref{connections});
    \item The low temperatures and weak vertical mixing meaning that photochemistry occurs in a high-pressure, low-temperature regime, and a low homopause permitting the formation of an extensive ionosphere (Section \ref{chem};
    \item The small-scale cloud phenomena, episodic outbursts, equatorial waves, and moist convective activity operating at shallow and accessible depths, potentially driven by CH$_4$ condensation and ortho/para-H$_2$ interconversion (Section \ref{meteorology};
    \item The stabilising effects of phase transitions implying layered tiers of circulation patterns and convection, potentially decoupled from one another as a function of depth (Sections \ref{circulation} and \ref{meteorology}).
\end{itemize}

The above list is not intended to be exhaustive, but provides numerous compelling reasons for future exploration of Uranus, both as a fascinating object in its own right, but also as one of our closest and best examples of a class of worlds that might be commonplace, being only slightly larger than the `sub-Neptunes' that currently dominate the census of exoplanets \citep{18fulton}.  Uranus and Neptune will be the last remaining class of Solar System planet to have a dedicated orbital explorer, and international efforts are under way to develop an ambitious mission to Uranus in the coming decades, combining an orbital tour with \textit{in situ} descent probes \citep{12arridge, 18mousis, 19hofstadter, 19fletcher_V2050, 19simon_probe, 20fletcher_philtrans}.  

Such a mission would be transformative for our understanding of Uranus, in the same way as Cassini, Galileo, and Juno transformed our understanding of the Gas Giants.  Key questions to be addressed by such a mission, supported by Earth-based multi-spectral remote sensing, are as follows:
\begin{itemize}
    \item What is the bulk composition of Uranus' interior, particularly the ice-to-rock ratio, elemental abundances (including noble gases) and isotopic ratios?  How well does the atmospheric composition represent the bulk, and how do water-rich oceans/mantles influence the observed composition?  These will provide crucial constraints on Uranus' formation and migration by revealing which reservoirs were available to the forming Uranus.
    \item What are the dynamical, meteorological, and chemical impacts of the negligible planetary luminosity, weak vertical mixing, and Uranus' extreme seasons, and how do atmospheric phenomena differ between Uranus and Neptune?
    \item What is the large-scale circulation of Uranus' atmosphere, how deep does it penetrate into the interior and is it coupled to interior motions, and do stable layers produce decoupled tiers of circulation and convection?
    \item What is the role of moist convection and precipitation in Uranus' hydrogen-dominated atmosphere, and how do updrafts overcome the static stability of the cloud bases?  
    \item Does Uranus really have no appreciable internal heat, or is this merely trapped, or time variable?  If so, what does this imply about Uranus' thermal evolution, and why does it differ from Neptune?
    \item What are the sources of energy responsible for heating the middle and upper atmosphere of Uranus, and how does the ionosphere couple the atmosphere to the external magnetosphere?
    \item How does Uranus' atmosphere evolve with time?  Is the banded structure relatively stable, or do the winds, cloud features, and albedo patterns shift over time?  How do vortices form, migrate, and interact in Uranus' atmosphere?
\end{itemize}

To date, progress in addressing these questions has been hampered by the extreme challenge of observing Uranus in the decades since Voyager.  Nevertheless, this review demonstrates how far we have come in developing our understanding of this unusual world, and whets our appetite for exciting new discoveries in the coming decades.

%% IMPORTANT! The old "\acknowledgment" command has be depreciated. It was
%% not robust enough to handle our new dual anonymous review requirements and
%% thus been replaced with the acknowledgment environment. If you try to 
%% compile with \acknowledgment you will get an error print to the screen
%% and in the compiled pdf.
\begin{acknowledgments}
Fletcher is supported by a European Research Council Consolidator Grant (under the European Union's Horizon 2020 research and innovation programme, grant agreement No 723890) at the University of Leicester.  I would like to extend my thanks to Michael Roman for providing the artwork for Fig. \ref{temp-clouds}, Ricardo Hueso for Fig. \ref{clouds}, to Julianne Moses for providing the model data plotted in Fig. \ref{cxhy}, to Glenn Orton for providing guidance for spectroscopy in Fig. \ref{ura_spectra}, and to Helmut Feuchtgruber and Martin Burgdorf for allowing me to show the ISO observations in Fig. \ref{ura_spectra}.

\end{acknowledgments}

%% To help institutions obtain information on the effectiveness of their 
%% telescopes the AAS Journals has created a group of keywords for telescope 
%% facilities.
%
%% Following the acknowledgments section, use the following syntax and the
%% \facility{} or \facilities{} macros to list the keywords of facilities used 
%% in the research for the paper.  Each keyword is check against the master 
%% list during copy editing.  Individual instruments can be provided in 
%% parentheses, after the keyword, but they are not verified.

%\vspace{5mm}
%\facilities{}

%% Similar to \facility{}, there is the optional \software command to allow 
%% authors a place to specify which programs were used during the creation of 
%% the manuscript. Authors should list each code and include either a
%% citation or url to the code inside ()s when available.

%\software{}

%% Appendix material should be preceded with a single \appendix command.
%% There should be a \section command for each appendix. Mark appendix
%% subsections with the same markup you use in the main body of the paper.

%% Each Appendix (indicated with \section) will be lettered A, B, C, etc.
%% The equation counter will reset when it encounters the \appendix
%% command and will number appendix equations (A1), (A2), etc. The
%% Figure and Table counter will not reset.

%\appendix

%\section{Appendix}

\bibliography{references.bib}{}
\bibliographystyle{elsarticle-harv}
%\bibliographystyle{apacite}
%% This command is needed to show the entire author+affiliation list when
%% the collaboration and author truncation commands are used.  It has to
%% go at the end of the manuscript.
%\allauthors

%% Include this line if you are using the \added, \replaced, \deleted
%% commands to see a summary list of all changes at the end of the article.
%\listofchanges

\end{document}